# An Anti-Turing Test: Reduced Variables for Social Network Friends' Recommendations

e-print

Iaakov Exman and Alex Krepch

Software Engineering Department
The Jerusalem College of Engineering - Azrieli
POB 3566, Jerusalem, 91035, Israel
`iaakov@jce.ac.il, alexkrepch@gmail.com`

Categories and subject Descriptors

H.3.3  [Information Search and Retrieval]: Retrieval models, Search Process {1998 ACM Classification}

Information Systems → Information Retrieval → Retrieval models and ranking
→ Novelty in information retrieval
Networks → Network types → Overlay and other logical network structures
→ Online social networks {CCS 2012}

**Abstract.** A routine activity of social networks' servers is to recommend candidate friends that one may know and stimulate addition of these people to one's contacts. An intriguing issue is how these recommendation lists are composed. This work investigates the main variables involved in the recommendation activity, in order to reproduce these lists including its time dependent characteristics. We propose relevant algorithms. Besides conventional approaches, such as friend-of-a-friend, two techniques of importance have not been emphasized in previous works: randomization and direct use of *interestingness* criteria. An automatic software tool to implement these techniques is proposed. Its architecture and implementation are discussed. After a preliminary analysis of actual data collected from social networks, the tool is used to simulate social network friends' recommendations.

# 1  Introduction

Social networks are very prominent in what is meant nowadays by software, being one of the most visible content transformers. Among other aspects, social networks continuously offer lists of suggested friend candidates. Friends' dynamics is one of the central activities of such networks.

In this paper – a revision and extension of the work published in [2] – we analyze data retrieved from recommendation lists of large public social networks, to obtain important factors influencing the generation of such lists and respective algorithms of relevance. In particular we pay attention to the order of suggested candidates within a given list and the variability of the lists' composition along the time.

Recommendations lists are very intriguing. One can hardly believe how some of the candidates appear there. In this work we do *not* focus on candidates known to the recommendation receiver. Our attention is directed to previously unknown candidates, and how they are possibly selected from a large social network database.

We coin these lists a sort of anti-Turing test, since we ask ourselves to what extent humans are stereotyped by a reduced number of variables. From the social networks' point of view people seem to look like artificially generated software constructs, rather than real people with a complex personality.

The ultimate test for the validity of our analysis is the ability to reproduce in a time-dependent fashion the general characteristics of the recommendation lists one receives. To this end a software tool was proposed and is being continuously developed.

## 1.1   Related Work

Here we present a concise review of related work.

First, we refer to recommendations and connections. Chen et al. [1] studied four recommendation algorithms:

1. *Content Matching* – closely related to finding documents of similar content;
2. *Content*-Plus-Link – adds to the content matching, the existence of a social link between the candidate and the recommendation receiver;
3. *Friend*-of-Friend – considers only social network structure;
4. *SONAR* – aggregates social relationship information from different public data sources (within IBM).

Their conclusion was twofold: algorithms based on social network information produced better-received recommendations and found more known contacts for users; algorithms using content similarity were stronger in discovering new friends.

Roth et al. [13] describe an *implicit social graph* and use it together with interaction-based affinity in suggesting friends. Huberman et al. [8] specifically analyze Twitter. They conclude that what really matters is a sparse and *hidden network of connections* underlying the declared set of friends and followers.

Golbeck and Hendler [5] investigated how trust information can be inferred from social network members not directly connected and integrated into applications, such as TrustMail, an email client. Tang et al. [14] deal with automatic labeling the "intensity" of social relationships, say "colleagues" or "friends".

Next, we refer to properties and actions on social networks.

Konstas and collaborators [10] deal with collaborative recommendation in social networks, using for instance, linear filtering.

Tools for various actions on social networks include the Referral Web by Kautz et al. [9]. Gross and Acquisti [6] refer to the problem of privacy in online social networks, in particular Facebook.

Finally, we refer to the notion of *anti-Turing test*. This expression has been used both in philosophical contexts and while discussing computer-human interfaces. For instance, Faith in his philosophical doctoral thesis [4] has a sub-section "Anti-Turing" of a chapter on "Intentionality: Insides", in which he asks: "How can we tell what is going on in someone's head?" obviously referring to Turing's test of artificial intelligence.

Laurel [11],[12] refers to the design of computer-human interfaces, suggesting that they should pass an anti-Turing test, to assure that humans are not confused to think that behind anthropomorphic interface agents there are real humans. Laurel has been cited a few times, e.g. by Guenter and Morrison [7].

In the remaining of the paper we introduce recommendation variables of importance (section 2), provide new kinds of relevant recommendation algorithms (section 3), overview the software architecture and implementation of our tool (section 4), describe a preliminary analysis of social networks data (section 5), and conclude with a discussion (section 6).

## 2   Recommendation Variables

A basic question of this work is: to what extent humans are adequately stereotyped by a reduced number of variables? In other words, is the human stereotype a kind of Anti-Turing test?

In order to try to answer this question, we first of all list potential variables of interest for friend recommendation.

Recommendation variables can be roughly classified into two types:
- *Unary contents* of the given person – either the candidate or the receiver of the recommendation;
- *Binary or multiple interactions* – among two or more members of the social network, either directly or indirectly.

The answer to the basic question will be empirically given by testing what is the minimal number of variables that will allow reasonable simulation of the generation of friends' recommendation lists by social networks.

### 2.1   Unary Content Variables

Unary content variables include among others, those related to profession and occupation:

- *Profession* – acquired in a certain institution; degree obtained;
- *Education institutions* – say high-schools, junior colleges or universities;
- *Employer* – company, public service or other organizations;

- *Occupation* – this may differ from the profession; rank in the organization;
- *Specific skills* – within the profession and/or occupation.

Other unary content variables may refer to:
- *Languages spoken* – local country or foreign languages;
- *Hobbies* – leisure activities;
- *Geographical location* – of residence and work; country, state, county, city;

Note that one could look at each one of the unary variables as sub-set collections or ranges. For example, universities could be classified into sub-sets – say ivy league.

### 2.2    Multiple Interaction Variables

Binary or multiple interaction variables can be direct, among people who know each other, such as:

- *Joint Publications* – co-authors of the same paper or book;
- *Exchanged Messages* – email, phone conversation, if data is available from providers;

Indirect interactions are possible also among people who are not mutual acquaintances:

- *Common Friends* – the person receiving the recommendation and the friend candidate have common friends; this touches the widely referred issue of *transitivity*;
- *Common Search topics* – again if data is available from providers of search engines.

## 3        Kinds of Relevant Recommendation Algorithms

Algorithms for recommendation list generation may involve semantic considerations – e.g. friend-of-a-friend – as well as abstract mathematical operations involving linear or non-linear filtering. Coefficients in linearly weighted sums express the relative importance of variables involved and should be usually normalized. Non-linear expressions may impart different orders of magnitudes to the variables' importance.

Besides the kinds of recommendation algorithms referred to in the related work (sub-section 1.1), we describe here two additional kinds: a- Interestingness; b- Degrees of randomness.

### 3.1    Interestingness

Interestingness according to Exman [3] is a function of both relevance and unexpectedness. The latter quantities can be themselves expressed in various forms

Anti-Turing – Friends Recommendations        Iaakov Exman & Alex Krepch

and also be combined in more than one way. The simplest way is just a multiplication:

$$\textbf{Interestingness = Relevance} \ast \textbf{Unexpectedness} \qquad (1)$$

One form to express relevance is by calculating the *match* of a candidate to contents defining a domain. The respective unexpectedness is calculated by a measure of *mismatch* to the domain. Together, with a normalization factor NormF, this is written as:

$$\textbf{Interest = Match * Mismatch / NormF} \qquad (2)$$

We advance the idea that interestingness for friend candidates, in which the domain is given by the recommendation receiver contents, is similar to that of any other content retrieval.

### 3.2 Degree of Randomness

Randomness may be applied within a sample of candidates at a given time stamp, or may be applied along the time axis.

Our observations show three possible behaviors of variables within a sample:

- *Constant* – very regular throughout the whole range;
- *Random* – according to a certain probability distribution;
- *Recognizable trend* – it is neither constant nor random; one can recognize definite functional trends, to be discussed below.

Recommendation lists may change along the time axis when a candidate is accepted as a new friend. It may also change just with time elapsing, even when no candidates were added since the previous visit to the member's page in the network.

## 4 The *RECOMM* Tool Architecture and Implementation

A tool called *RECOMM* has been gradually developed to test the hypotheses that we propose. Here its architecture and implementation are concisely described.

Fig. 1 shows the following internal modules:

- *Randomize inputs* – to select random candidates to add to the previous recommendation list; to determine their order in the recommendation list; to assign values to chosen variables;
- *Interestingness* – to increase the chances of candidates in detriment of others, based on their potential interest to the recommendation receiver;
- *Calculate Recommendation* – combining the above algorithms with the friend-of-a-friend alrgorithm;
- *Sorting & Threshold* – according to recommendation grades;

Anti-Turing – Friends Recommendations    Iaakov Exman & Alex Krepch

- *Decoration* – add a picture of the candidate and some possible additional features for display.

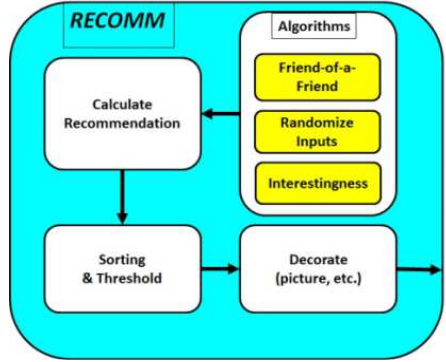

**Fig. 1.** *RECOMM* Architecture –modules displayed as upper states in the system statechart.

*RECOMM* has been implemented in C#. The respective class diagram is seen in Fig. 2. It has five important classes, with some of their fields and methods displayed.

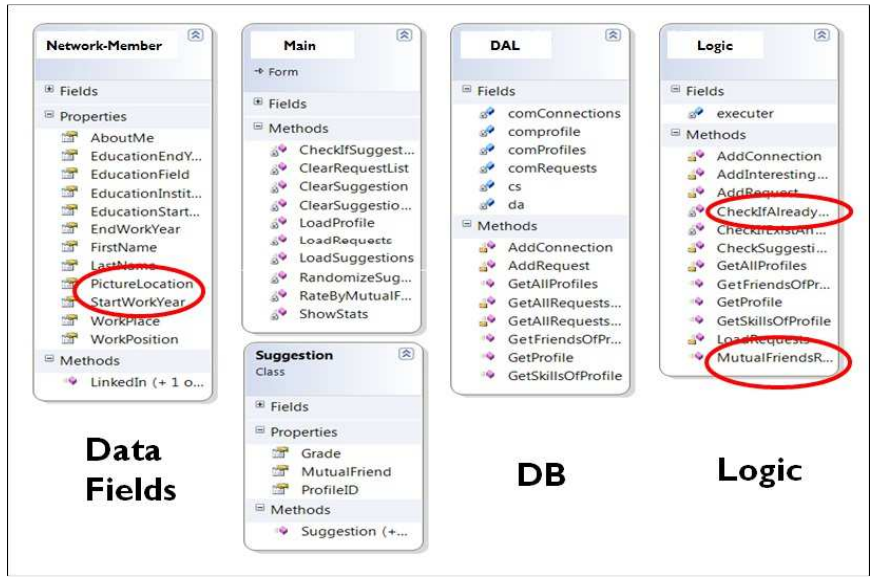

**Fig. 2.** *RECOMM* class diagram – From left to right one can see the following classes: a- the Network-Member receiving the suggestion, with a red ellipse marking the location of his picture; b- Main class; c- specific Suggestion class; d- DAL, a database related class; e- the Logic class, displaying within red ellipses, methods such as checking whether the Network-Member and the suggestion have MutualFriends.

## 5  Preliminary Data Analysis of Social Networks

### 5.1   Data Collected

We have collected data from pages of members[1] of large social networks, to make a preliminary analysis, pointing to novel recommendation approaches.

For each member page and time stamp, we collected samples containing the first 50 recommendations, with the values of the available variables. Conclusions were inferred from the analysis of values within and among samples along time, for given social networks.

A sample form to collect data on friend candidates is seen in Fig. 3. It corresponds to a given time stamp (data & time) and it has 50 records, of which only 3 are shown.

| Social Network – Friend Candidates - **FORM** ||||||
|---|---|---|---|---|---|
| Date: 12/07/2013 ||||||
| Time: 22:00 ||||||
| # | Name | Degree | Shared connections | Known from | Position | Comments |
| 1 | John Doe | 2 | 21 | studies | Computer Software Professional | |
| 2 | Richard Roe | 2 | 2 | | College Student | |
| 3 | Joe Blogs | 3 | - | brother | Web Market Expert | |

**Fig. 3.** Sample form to collect Friend Candidates date – Each row refers to a friend candidate. Each form has 50 rows. Columns refer to candidate properties, to be edited into variable values.

Collected data are presented next under the rubric of the conclusions inferred, in order to provide support for the conclusions.

### 5.2    Data Distribution Trends

For LinkedIN, variables with *constant behavior* include:

- the network *degree* – the distance between the candidate and a friend of the receiver – is almost always "2$^{nd}$", i.e. the receiver has a friend directly connected to the candidate, implying a systematic "friend-of-a-friend" policy;
- the number of *shared connections* – in some samples may be almost constant with few exceptions;
- whether the *candidate is known* to the recommendation receiver – a Boolean variable; also in some samples, the candidates are almost always "unknown" to the receiver, again with few exceptions.

---

[1] Data was collected with the active help or consent of the respective page owners: the paper authors and their friends.

Anti-Turing – Friends Recommendations          Iaakov Exman & Alex Krepch

Still for LinkedIN, a localized recognizable trend seems to be:

- the first two candidates – may be "known" to the receiver. This is a sort of stimulus to accept the suggested candidates, and it certainly differs from the constancy of the above mentioned variables.

The other variables for these candidates may be rather random. Thus, the network degree may be different from "$2^{nd}$" and the number of shared connections may be any value. Besides the first candidates, other randomly placed candidates – in varying fractions of the distribution – may also be "known".

For Facebook, a recognizable trend for shared connections is:

- a unimodal distribution, whose peak is at low values. A plot of the histogram of shared connections for a given sample is seen in Fig. 4.

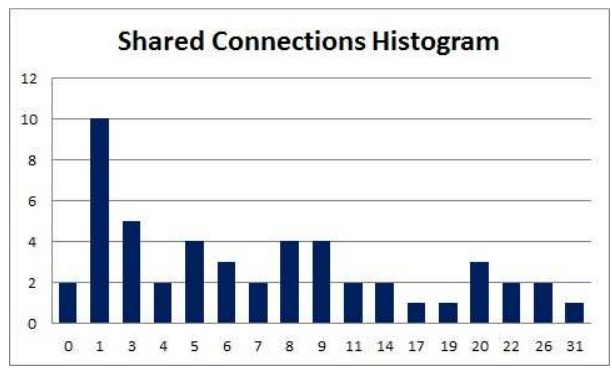

**Fig. 4.** Shared connections histogram – these are shown for a Facebook sample of suggested friend candidates. One can see the relative diversity of shared connection values in the horizontal axis. The unimodal peak is at the value 1 shared connection, with 10 candidates displaying this value.

It can be said that despite the diversity of shared connection values, as shown in Fig. 4., many of them with possibly known friends, the peak at the low value indicates a recommendation policy emphasizing *novelty*. This means that most of the recommended suggestions have a majority of low shared connections, stimulating the establishment of novel connections, to amplify personal networks. This seems to be a consistent policy of social networks in general.

Concerning changes of distribution along the time axis, the changes of recommendation lists within a short time (say a few hours) may be very dramatic. Thus the important variable in this respect is not absolute time, but the fact that it is a new visit to the member's page.

### 5.3   Explanation by Interestingness

There are variables in which there is a non-localized recognizable trend, which demands some deeper explanation.

For LinkedIN, one such variable is:

- the geographical location of the workplace – If one takes the locations' distribution provided by the social network for the friends of the given member (the recommendation receiver), one finds that it is a very different distribution from that of the candidates in the recommendation list.

For instance, for a given sample the member friends are located in 6 countries, with the big majority in the country of the member himself. The candidates are located in 4 countries, with the majority still in the country of the member. But the second country in terms of candidates has a very significant increase relatively disproportional to the member's friends – say about 30% instead of less than 10%.

A tentative explanation is as follows. The *unexpected* increase of this variable is due to the potential *interestingness* of this second country. On the one hand, it is *relevant*, i.e. a mainstream country in terms of the member friends, hinting to a potential increase. On the other hand it is *unexpected*, i.e. its contribution to the distribution is unexpected given some of its characteristics, say country size, distance from the member's country, or international relations.

## 6   Discussion

A framework was proposed for generating recommendations of friendship candidates in a given social network. The framework contains the important variables for given social networks, recommendation algorithms and a set of controls to output a recommendation list.

The framework has been implemented in the *RECOMM* simulator tool, to test the hypothesis concerning reduced number of variables to characterize friends recommendation lists by social networks.

### 6.1   Validation

Validation of *RECOMM* output has been performed against actual recommendation lists of specific social networks, e.g. LinkedIN and Facebook, for data obtained from members of these networks.

Preliminary collection of data has been performed and analyzed. The simulator does not yet reproduce faithfully actual recommendation lists, due to lack of precision of data gathered.

Preliminary conclusions include:

a-  Apparently, friends' recommendation lists are generated by different techniques for distinct social networks;
b-  Friend-of-a-friend is clearly important; it may be weighted by desired coefficients, say workplace;

   c-   Randomization and interestingness seem to be relevant and promising algorithms;

### 6.2 Future Work

After full development of the *RECOMM* tool, it will be extensively used to test the hypotheses advanced about the relative importance of the above mentioned variables and algorithms for specific social networks.

*RECOMM* output of recommendation lists will be statistically more precisely characterized for similarity to the collected data.

An interesting issue is the degree of generality of the chosen variables and algorithms for diverse social networks, i.e. to what extent the tool will need fine tuning to apply it to each different network.

### 6.3 Main Contribution

The main contribution of this work is the recognition of the importance of randomization and interestingness, given the reduced number of variables to generate friends' recommendation lists.